\documentclass[reprint,aps,prapplied,amssymb,lipsum,twocolumn,showpacs,floatfix,superscriptaddress,nobalancelastpage,footinbib]{revtex4-1}

\usepackage{graphicx}
\usepackage{multirow}
\usepackage{amssymb}
\usepackage{amsmath}
\usepackage{color}
\usepackage{url}
\usepackage{epstopdf}

\usepackage[breaklinks, urlcolor=blue, hyperindex, colorlinks, bookmarks=true, citecolor=blue,]{hyperref}

\usepackage{microtype}

\graphicspath{{Figures/}}
\begin{document}

\title{Coupling a Superconducting Qubit to a Left-Handed Metamaterial Resonator}

\author{S. Indrajeet}
\affiliation{Department of Physics, Syracuse University, Syracuse, NY 13244-1130}

\author{H. Wang}
\altaffiliation{Present address: NIST Quantum Electromagnetics Division,University of Colorado Boulder, Boulder, CO 80305}
\affiliation{Department of Physics, Syracuse University, Syracuse, NY 13244-1130}

\author{M.D. Hutchings}
\altaffiliation{Present address: SeeQC, Inc., Suite 141, 175 Clearbrook Road, Elmsford, NY 10523, USA}
\affiliation{Department of Physics, Syracuse University, Syracuse, NY 13244-1130}

\author{B.G. Taketani }
\affiliation{Departamento de F\'{i}sica, Universidade Federal de Santa Catarina, 88040-900 Florian\'{o}polis, SC, Brazil}

\author{Frank K. Wilhelm }
\altaffiliation{Present address: Institute for Quantum Computing Analytics (PGI 12), Research Center J\"{u}lich, 52425 J\"{u}lich, Germany}
\affiliation{Theoretical Physics, Saarland University, Campus, 66123 Saarbr\"{u}cken, Germany}

\author{M.D. LaHaye}
\altaffiliation{Present address:  United States Air Force Research Laboratory, Rome, New York, 13441 USA.}
\affiliation{Department of Physics, Syracuse University, Syracuse, NY 13244-1130}

\author{B.L.T. Plourde}
\email[]{bplourde@syr.edu}
\affiliation{Department of Physics, Syracuse University, Syracuse, NY 13244-1130}

\date{\today}

%\pacs{}

\begin{abstract} 
Metamaterial resonant structures made from arrays of superconducting lumped circuit elements can exhibit microwave mode spectra with left-handed dispersion for the standing-wave resonances, resulting in a high density of modes in the same frequency range where superconducting qubits are typically operated, as well as a bandgap at lower frequencies that extends down to dc. Using this  regime for multi-mode circuit quantum electrodynamics, we have performed a series of measurements of such a superconducting metamaterial resonator coupled to a flux-tunable transmon qubit. Through microwave measurements of the metamaterial, we have observed the coupling of the qubit to each of the modes that it passes through. Using a separate readout resonator, we have probed the qubit dispersively and characterized the qubit energy relaxation as a function of frequency, which is strongly affected by the Purcell effect in the presence of the dense mode spectrum. Additionally, we have investigated the ac Stark shift of the qubit as the photon number in the various metamaterial modes is varied. Through numerical simulations, we have explored designs based on this scheme with enhanced coupling so that the coupling energy between the qubit and the metamaterial modes can exceed the intermode spacing in future devices with achievable circuit parameters. The ability to tailor the dense mode spectrum through the choice of circuit parameters and manipulate the photonic state of the metamaterial through interactions with qubits makes this a promising platform for analog quantum simulation with microwave photons and quantum memories.

\end{abstract}

\maketitle

\begin{figure}[b]
\centering
\includegraphics[width=3.35in]{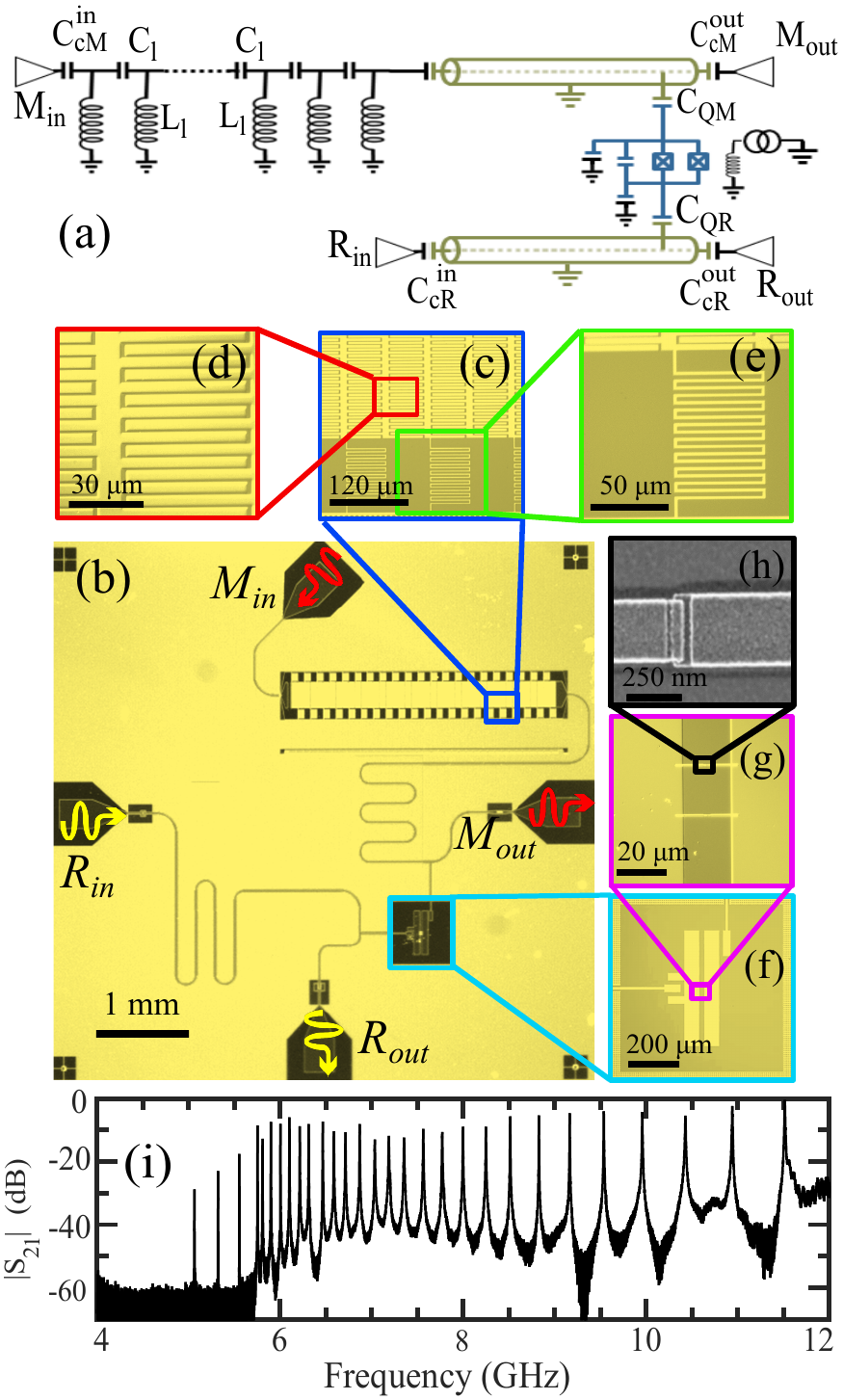}
  \caption{(a) Circuit schematic of device. (b-g) Optical micrographs of metamaterial resonator device: (b) zoomed-out image of entire chip, (c) section of metamaterial resonator containing several unit cells of inductors and capacitors, (d) detail of interdigitated capacitor, (e) detail of meander-line inductor, (f) transmon qubit with coupling capacitors to metamaterial resonator and readout resonator,(g) SQUID loop of the transmon, (h) Scanning electron microscope (SEM) image of one of the transmon junctions. (i) Transmission spectrum of metamaterial resonator measured through ports $M_{in}$ and $M_{out}$. 
\label{fig:meta-schem}}
\end{figure}

\section{Introduction}
\label{sec:Introduction}

Artificial atoms formed from Josephson junction-based devices combined with linear resonant circuits are capable of reaching the strong coupling regime, where the coupling strength between the qubit and resonant microwave mode is greater than the relevant linewidths \cite{Schuster2007}. 
The field of circuit quantum electrodynamics (cQED) \cite{Blais2004} has developed into one of the leading architectures for implementing scalable quantum processors \cite{Arute2019} while also providing a  regime for exploring the light-matter interaction in microwave quantum optics with tailored artificial atoms \cite{Blais2020}.
For systems with multiple modes coupled to a qubit, there is the possibility of analog quantum simulation with photons in the different modes, allowing for the realization of a quantum model in a controlled platform \cite{Cirac2012}. As one example, the light-matter interaction Hamiltonian at large coupling strengths lends itself to the realization of the spin-boson model  \cite{Egger13, messinger2018left, Leppakangas18,Forn-Diaz2017}, a paradigmatic model of quantum dissipation and quantum phase transitions. Multi-mode cQED can also be used for studying qubit dynamics in the vicinity of photonic bandgaps \cite{liu2017quantum, mirhosseini2018}, analogous to experiments using real atoms and photonic crystals \cite{Hood2016}. Quantum random access memories are yet another potential application for qubits coupled to multiple modes \cite{naik2017random}.

 A variety of routes for achieving multi-mode cQED have been studied, including continuous transmission-line resonators \cite{Sundaresan15} and metamaterials made from lumped circuit elements \cite{mirhosseini2018}  and Josephson junctions \cite{Kuzmin2019,Martinez2019,Hutter2011}, as well as qubits coupled to acoustic and electromechanical systems \cite{Moores18, Han2016}. Lumped-element metamaterial transmission lines can be configured to tailor the band structure and thus influence the wave speed and bandgaps in the system. Possible wave dispersion properties include a  left-handed dispersion relation where the mode frequency is a falling function of the wavenumber \cite{Eleftheriades2002, caloz2004tl}. Superconducting metamaterials can thus provide a low-loss system with an effective negative index of refraction \cite{Jung2014multi,jung2014progress}. When implemented in a cQED architecture \cite{Wang2019}, left-handed transmission line (LHTL) resonators provide a pathway for implementing a dense microwave mode spectrum above a low-frequency bandgap that extends down to zero frequency.

In this manuscript, we demonstrate the strong coupling of a superconducting transmon qubit  \cite{Koch2007} to such a LHTL resonator. Our circuit design allows us to measure the qubit state and perform coherent qubit manipulation while separately or simultaneously driving the metamaterial resonances. We are thus able to characterize the qubit lifetime while tuning its transition frequency through the dense metamaterial mode spectrum. In addition, we can probe the Stark shift of the qubit transition due to photons in the metamaterial modes through spectroscopic measurements while driving different system resonances. We also use numerical simulations to investigate future device designs with achievable circuit parameters that can reach the  regime with coupling strengths between the qubit and metamaterial modes that are larger than the intermode spacings. Our scheme thus demonstrates an approach to multimode cQED that provides a platform that can be used for quantum simulations with microwave photons or for dense quantum memories.

\section{Device Design and Experimental Configuration}
\label{Device Design}
%BP edits -- 9/22/19

Conventional cQED architectures involve a resonator that has a standard right-handed dispersion, where the mode frequency is an increasing function of wavenumber. Such resonators are typically formed from a distributed transmission-line segment, such as a coplanar waveguide (CPW). In principle, lumped-element implementations of the transmission-line resonator are also possible with a 1D array of series inductors and shunt capacitors to ground \cite{Pozar-book}. By contrast, building a LHTL must be done with a lumped-element metamaterial approach \cite{Eleftheriades2002}. Swapping the circuit-element positions so that the array contains series capacitors $C_l$ with inductors $L_l$ to ground [Fig.~\ref{fig:meta-schem}(a)] results in a low-frequency bandgap from dc up to an infrared cutoff frequency $\omega_{IR}=1/2\sqrt{L_l C_l}$. Frequencies just above $\omega_{IR}$ correspond to the {\it shortest} wavelengths in the system, while the wavelength {\it increases} for progressively higher frequencies -- a characteristic signature of left-handedness. A resonator formed from a finite-length LHTL with coupling capacitors at either end will exhibit a dense spectrum of modes just above $\omega_{IR}$ where the band is nearly flat with respect to wavenumber \cite{Wang2019}.

Coupling of transmon qubits to resonant circuits is commonly achieved through a weak capacitance to a portion of the resonator near a voltage antinode of a particular mode. For a multimode system, the standing-wave pattern   will in general be different for each mode. Thus, for a given qubit placement, strong coupling is only possible to the modes with large amplitudes near the qubit. For our metamaterial resonator, by building a hybrid system with one end of the LHTL connected to a right-handed transmission line (RHTL), the modes in the frequency range accessible by a typical transmon will have a short wavelength in the LH section with a much longer wavelength in the RH portion. The mode spectrum for such a hybrid LHTL-RHTL resonator still exhibits peaks near the LHTL resonances, with a slow modulation from the standing wave in the RH portion. Thus, by placing the qubit near the end of the RHTL, close to the antinodes of the low-frequency modes, the qubit can couple to all of the modes in its frequency range \cite{Egger13}.

Figure~\ref{fig:meta-schem}(a) contains a schematic of our device. The input (output) coupling capacitor $C_{cM}^{in}$ ($C_{cM}^{out}$) near the end of the LHTL (RHTL) section define the resonant modes of the hybrid transmission line structure. The transmon qubit follows the design in Ref. \cite{Chow2015spie} and contains a split junction allowing for the qubit transition frequency to be tuned with an external magnetic flux, which is provided by a wirewound coil above the chip. Capacitor $C_{QM}$ couples the qubit near the output end of the RHTL portion of the hybrid metamaterial; a second capacitor, $C_{QR}$, couples the qubit to a separate half-wave RHTL resonator for conventional dispersive readout of the qubit state \cite{Blais2004}.

\begin{figure}[htb]
\centering
\includegraphics[width=3.35in]{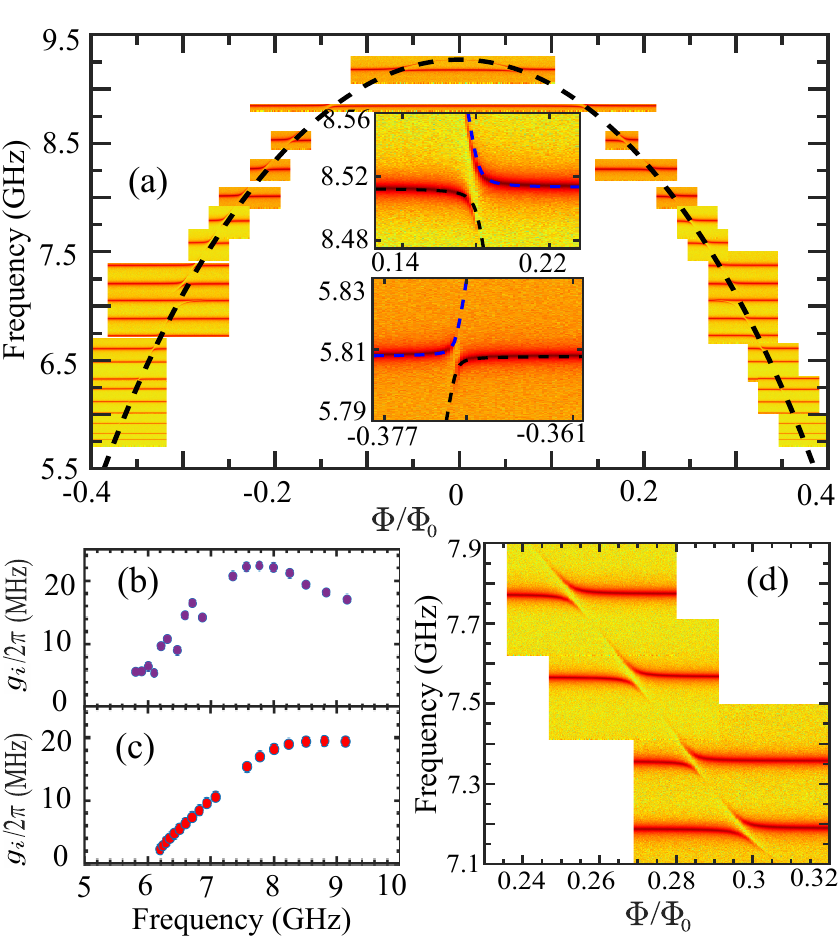}
  \caption{ (a) Close-up density plot of magnitude of transmission through metamaterial resonator vs. qubit flux bias, showing vacuum Rabi splittngs for all of the modes that the qubit tunes through. Dashed line indicates flux tuning of bare transmon energy band from fits to splitting location. (Inset) closeup of splittings and fits for 5.81~GHz and 8.52~GHz modes. (b) Extracted coupling strength between qubit and each metamaterial mode vs. frequency. (c) Simulated coupling strengths vs. frequency using AWR Microwave Office and circuit model approach described in text. (d) Detail of splittings for four adjacent modes between 7.1 and 7.9~GHz.    
\label{fig:vacuumrabi}}
\end{figure}

The LHTL portion of the hybrid metamaterial resonator is fabricated from an array  of 42 unit cells of series interdigitated capacitors with 29 pairs of 4-$\mu$m wide/52-$\mu$m long fingers. Between each pair of capacitors is a meander-line inductor with 9 turns of 2-$\mu$m wide traces; the inductors are arranged in a staggered pattern with alternating inductors connected to the ground plane on either side of the LHTL, as in Ref.~\cite{Wang2019}. The input coupling capacitor  to the LHTL portion $C_{cM}^{in}$ is formed with a 4.9-$\mu$m wide gap to the input feedline. The RHTL portion of the hybrid metamaterial resonator (readout resonator) is formed from a 5-mm (7.7-mm) long  CPW with a 10-$\mu$m wide center conductor and a 6-$\mu$m wide gap to the ground plane on either side. These parameters  allow us to match the impedance (about 50 Ohms)   between  the LHTL and RHTL segments. The output end of the RHTL consists of an interdigitated capacitor $C_{cM}^{in}$  for coupling to the output feedline.  (Table \ref{table:meta}). 
%The RHTL portion and the readout resonator are formed from  CPW structures.
 The various circuit elements are patterned photolithographically with an etch process to define the structures in a Nb thin film on Si [Fig.~\ref{fig:meta-schem}(b-g)], with the exception of the Al-AlOx-Al Josephson junctions, which are defined with a shadow-evaporation process [Fig.~\ref{fig:meta-schem}(h)]. Further details of the fabrication and device parameters are contained in  Appendices \ref{app:fab} and \ref{app:para}.

We cool the device on a dilution refrigerator with a base temperature around 25~mK. With a vector network analyzer and microwave leads coupled to ports $M_{in}$ and $M_{out}$ [Fig.~\ref{fig:meta-schem}(b)], we  measure the microwave transmission $S_{21}(f)$ and probe the modes of the LHTL-RHTL resonator [Fig.~\ref{fig:meta-schem}(h)]. The transmission below $\omega_{IR}/2\pi \approx 5\,{\rm GHz}$ is vanishingly small, followed by a dense spectrum of resonance peaks for higher frequencies. As described in Ref.~\cite{Wang2019}, the staggered inductor configuration of our LHTL results in the band being not quite flat at $\omega_{IR}$ so that the densest region of the spectrum is roughly 1~GHz higher than $\omega_{IR}/2\pi$.

\section{Probing metamaterial modes coupled to qubit}
\label{sec:Vacuum Rabi Spilitings}

Initially we probe the qubit indirectly through transmission measurements of the hybrid metamaterial, rather than using the separate readout resonator. By varying the flux bias to tune the qubit transition frequency, we are able to observe the influence of the qubit on each mode through which it passes. The qubit has a maximum transition frequency of 9.25~GHz at an integer flux quantum ($\Phi_0 \equiv h/2e$, where $h$ is Planck's constant and $e$ is the electron charge) bias, thus allowing it to tune through all 21 modes below this point. 

For each of these modes, when the flux bias  results in the bare qubit frequency approaching a resonance, we observe a vacuum Rabi splitting \cite{Wallraff2004} in the microwave transmission where the qubit hybridizes with the photonic state of the metamaterial resonant mode [Fig.~\ref{fig:vacuumrabi}(a)]. Since each of these splittings is larger than the linewidth of the particular mode, this corresponds to the strong coupling regime of cQED \cite{schuster2007circuit}. We can fit the positions of these various splittings to the conventional transmon flux-modulation dependence \cite{Koch2007} while varying the maximum Josephson energy $E_{J0}$ to obtain the black dashed curve drawn on  Fig.~\ref{fig:vacuumrabi}(a). The extracted value for $E_{J0}$ is consistent with the fabricated junction parameters.

Since the splittings are smaller than the intermode spacings over the range of our measurements, we can make an approximation and treat each mode individually for extracting the coupling strength $g_i$ for the qubit to each of the metamaterial modes. In this way, we fit each vacuum Rabi splitting to the Hamiltonian for one resonator mode coupled to a transmon. Further details on this fitting, including a discussion of the validity of treating each mode separately, are included in  Appendix \ref{app:extraction}. The insets to Fig.~\ref{fig:vacuumrabi}(a) contain two examples of these fits to the splittings while Fig.~\ref{fig:vacuumrabi}(b) shows a plot of the extracted $g_i$ values to each mode. Initially $g_i/2\pi$ increases with frequency above $\omega_{IR}$ up to a maximum of 22~MHz for the mode near 7.8~GHz; this is followed by a gradual decrease in $g_i$ up to the maximum qubit transition frequency. This non-monotonic variation of $g_i$ with frequency is due to the standing-wave structure in the RHTL portion of our hybrid metamaterial resonator.

For modeling $g_i$ and comparing with the measurements, we use AWR Microwave Office \cite{AWR}. We can simulate the splittings in the spectrum semi-classically by approximating the qubit as a tunable $LC$ oscillator coupled to a hybrid metamaterial resonator with the parameters of our device. Details are discussed in Appendix~\ref{app:gawr}. The simulated $g_i$ frequency dependence plotted in Fig.~\ref{fig:vacuumrabi}(c) is in reasonable agreement with our measured coupling strengths, although the decrease in $g_i$ for the highest qubit frequencies is not quite as strong. The circuit simulation allows us to explore an artificial qubit with a much higher maximum frequency so that we can study $g_i$ further along the metamaterial resonance spectrum. The resulting plot of $g_i$ vs. frequency in the Appendix~\ref{app:gawr} (Fig. \ref{fig:awrg20GHz}) exhibits multiple dips, where the reduction that we observe in our experiment around $\sim 8-9\,{\rm GHz}$ is the first dip, due to the standing-wave structure in the RHTL portion.

\begin{figure}[tb]
\centering
\includegraphics[width=3.35in]{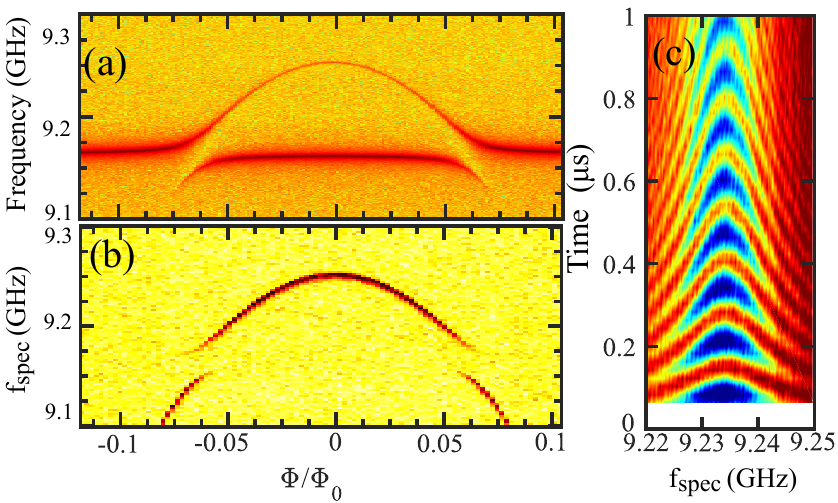}
  \caption{ (a) Close-up density plot of magnitude of transmission through metamaterial resonator vs. qubit flux bias in vicinity of upper sweetspot. (b) Spectroscopy of qubit 0-1 transition vs. flux bias near the upper sweetspot measured using readout resonator. (c) Rabi spectroscopy of qubit for flux bias near upper sweet spot, around 9.235~GHz, measured using readout resonator.
\label{fig:spectroscopy}}
\end{figure}
In the region around 7.5~GHz, where the modes are relatively close together and the $g_i$ values are near the maximum, the upper branch of the splitting for one mode comes close to touching the lower branch of the splitting for the next higher mode [Fig.~\ref{fig:vacuumrabi}(d)]. Nonetheless, even the maximum $g_i$ remains smaller than the minimum mode spacing, thus, the system is not quite in the multimode strong coupling regime, where the qubit would be able to couple strongly to multiple modes simultaneously. For future devices based on this architecture, it should be possible to increase $g_i$ by increasing the coupling capacitance
to the qubit, increasing the impedance of the metamaterial, and optimizing the length of  RHTL portion of the resonator. In addition to increasing $g_i$, it would also be helpful to reduce the spacing between modes, which could be achieved by either increasing the number of unit cells or through appropriate engineering of the parasitic reactances. Details are discussed in Appendix~\ref{app:incg}.

\label{sec:lhlmeasurements}

\begin{figure}[htb]
\centering
\includegraphics[width=3.35in]{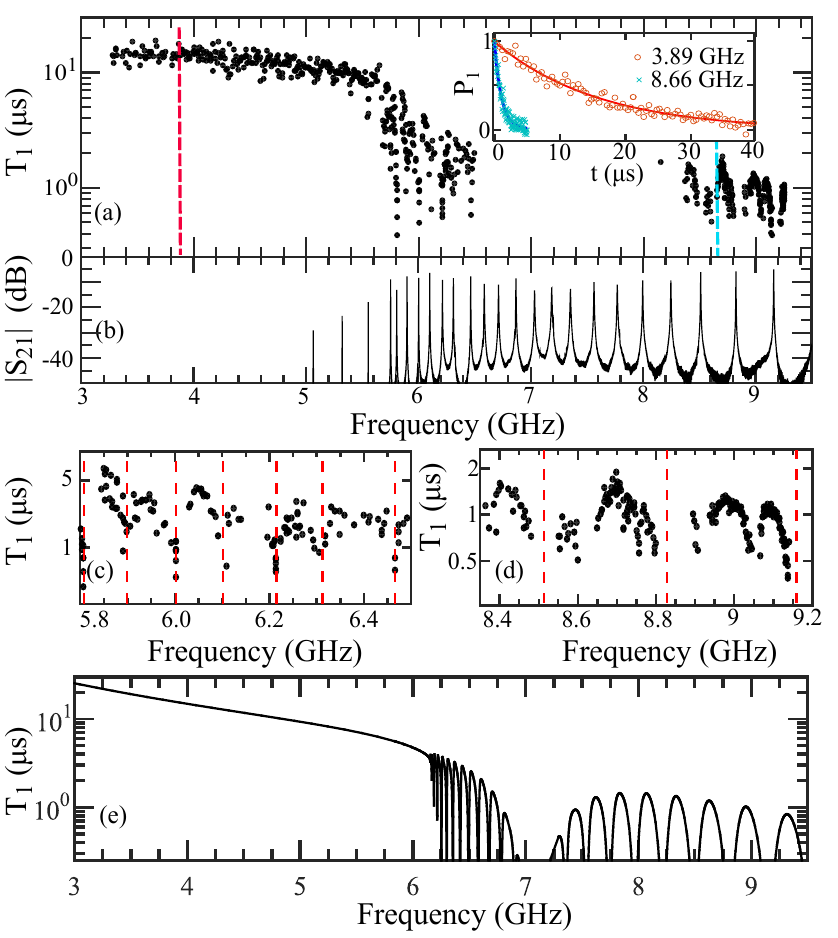}
  \caption{(a) $T_1$ vs. frequency measured over span of 6~GHz. (Inset) plots of two example measurements of qubit relaxation measured for a bias point below $f_{IR}$ of the metamaterial and a bias point in between two modes at higher frequencies. (b) Transmission spectrum of metamaterial resonator for comparison with structure in $T_1(f)$ plot. (c, d) zoomed-in $T_1(f)$ plots of data from (a) with red vertical dashed lines indicating location of metamaterial resonator modes. (e) Calculated $T_1(f)$ from multi-mode Purcell loss simulation of qubit coupled to metamaterial resonator.
\label{fig:T1}}
\end{figure}

\section{Purcell Losses in a Multimode Environment}

\label{sec:Qubit Relaxation}

In addition to probing the interactions of the qubit and metamaterial modes through direct measurements of the metamaterial resonator, we are also able to read out the qubit with the separate readout resonator. By monitoring the transmission between ports $R_{in}$ and $R_{out}$ [Fig.~\ref{fig:meta-schem}(b)] with heterodyne detection, we perform conventional dispersive measurement of the qubit state.  Figure~\ref{fig:spectroscopy}(a) is a plot of the vacuum Rabi splittings for the highest frequency mode that the qubit passes through, observed in the microwave transmission through the metamaterial resonator between ports $M_{in}$ and $M_{out}$ with network analyzer measurements, as in the previous section. In Fig.~\ref{fig:spectroscopy}(b), we use ports $R_{in}$ and $R_{out}$ to measure qubit spectroscopy in the same region of flux bias and frequency by sending a spectroscopy probe pulse at frequency $f_{spec}$, followed by a readout resonator pulse to detect the qubit-state-dependent dispersive shift. The two curves are quite similar, with comparable splittings when the bare qubit frequency passes through the 9.15~GHz mode.
%Compare with splittings obtained from vacuum Rabi measurements of other modes...?

We can also use the separate readout resonator for conventional coherent manipulation of the qubit state. Figure ~\ref{fig:spectroscopy}(c) shows a Rabi spectroscopy measurement of the qubit near its maximum transition frequency with a Rabi pulse of variable duration and frequency driven to the readout resonator, again followed by a resonator readout pulse. Thus, the qubit remains coherent in this frequency region, despite being biased in the middle of the complex resonance spectrum produced by the metamaterial resonator. A $\pi$-pulse tuned up from such a measurement allows us to excite the qubit and characterize its lifetime as it relaxes back to the ground state.

By stepping through the qubit flux bias and tuning up a $\pi$-pulse at each point, we are able to map out the qubit $T_1(f)$ in the structured environment of our hybrid metamaterial resonator [Fig.~\ref{fig:T1}(a)]. With the qubit biased below $\omega_{IR}/2\pi$, the qubit has a reasonably long lifetime with $T_1$ ranging between $10-19\,\mu{\rm s}$, with a gradual decrease for larger frequencies. For frequencies near $\omega_{IR}/2\pi$ and slightly below 6~GHz, $T_1$ begins to drop significantly, although notably the lowest three metamaterial modes, which are particularly high Q and low transmission, do not strongly influence $T_1$ [Fig.~\ref{fig:T1}(b)]. Beyond the fourth lowest mode, $T_1(f)$ is characterized by a series of sharp dips to sub-$\mu{\rm s}$ levels when the bare qubit frequency matches the various metamaterial resonances; there is a partial recovery of $T_1$ in between the modes [Fig.~\ref{fig:T1}(c, d)]. The gap in Fig.~\ref{fig:T1}(a) around 7-8~GHz is a result of the  strong coupling to the readout resonator, which makes it difficult to address the qubit directly.

The complex frequency dependence of the qubit lifetime that we observe is characteristic of Purcell loss for a qubit coupled to a series of lossy resonant modes \cite{Houck2008}. We note that the quality factors of the metamaterial modes in our device are entirely dominated by coupling losses to external circuitry. In this case, we chose $C_{cM}^{in/out}$ to be rather large to make it feasible to observe vacuum Rabi splittings in transmission measurements through the metamaterial while tuning the qubit frequency. For future devices, these coupling capacitances could be made significantly smaller if one were focused instead on probing the metamaterial modes via the qubit, rather than by direct transmission through the metamaterial. In this case the mode quality factors would be significantly higher and the Purcell losses much smaller, particularly when the qubit is tuned in between the modes.

 Following the approach outlined in Ref.~\cite{Houck2008}, we model the multi-mode Purcell effect as $T_1^{Purcell}(f) = C/{\rm Re}[Y(f)]$, where $C$ is the qubit shunt capacitance and $Y(f)$ is the frequency-dependent complex admittance of the qubit environment. We obtain $Y(f)$ using the circuit model of the impedance of a LHTL that we derived previously in Ref.~\cite{Wang2019} combined with the standard expression for a RHTL. Additionally we include a non-Purcell term $T_1^{non-Purcell}(f) \propto 1/f$ to model the effects of dielectric loss with a frequency-independent loss tangent that is typically observed in frequency-tunable transmons \cite{Barends13, Hutchings17}. Further details of the $T_1(f)$ modeling are described in Appendix~\ref{app:purcell}. The resulting $T_1(f)$ dependence from our model plotted in Fig.~\ref{fig:T1}(e) qualitatively follows our measurements, although we do not quantitatively reproduce the locations of the $T_1$ dips due to the difficulty of capturing the experimental metamaterial spectrum in our circuit model, particularly near $\omega_{IR}/2\pi$, without accounting for the effects of the staggered inductor configuration and non-ideal grounding of the metamaterial chip, as described in Ref.~\cite{Wang2019}. Nonetheless, the Purcell modeling provides a route for describing the frequency-dependent lifetime for a qubit coupled to a complex metamaterial, including the relatively long $T_1$ values that can be attained in the bandgap of the metamaterial spectrum.

\verb||

\section{Stark Shift Measurements}
\label{sec:Stark Shift Measurements}

\begin{figure}[!htb]
\centering
\includegraphics[width=3.35in]{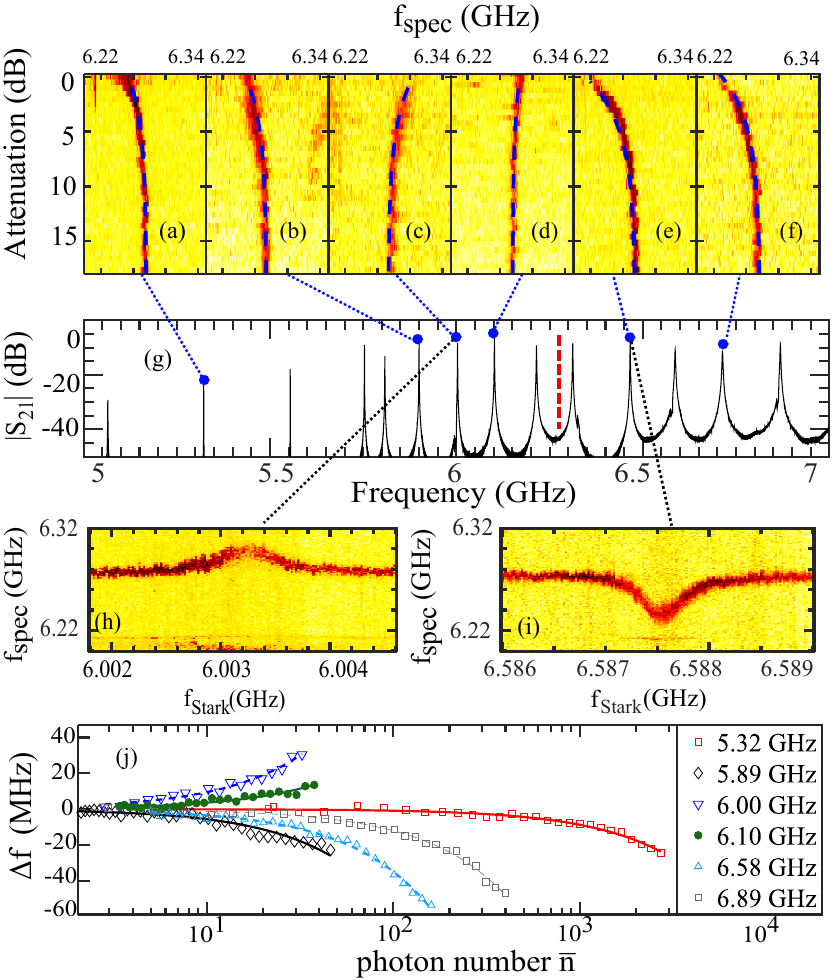}
  \caption{ (a-f) Stark shifts of qubit transition through driving of 6 different metamaterial modes at a range of microwave powers for a flux bias corresponding to the unshifted qubit transition at 6.275~GHz; Stark shift theoretical curves shown by blue dashed lines.  For the 6 plots, 0~dB attenuation corresponds to a power at the chip of (a) -86~dBm, (b) -126~dBm, (c) -128~dBm, (d) -123~dBm, (e) -121~dBm, and (f) -113~dBm. (g)  $|S_{21}(f)|$ measured through metamaterial to indicate the 6 modes driven in (a-f) (blue dots and dotted lines) and the bias point of the qubit (red dashed line). Plots of qubit spectroscopy frequency vs. Stark drive frequency for fixed power for (h) mode near 6.003~GHz at -129~dBm, (i) mode near 6.588~GHz at -122~dBm; metamaterial mode near 6.22~GHz visible as a faint, sharp line near bottom of plots. (j) Extracted Stark shifts in qubit transtion frequency from (a-f) plotted vs. mean photon number for each of the 6 metamaterial modes, as described in text.
\label{fig:Starkshifts}}
\end{figure}

\begin{figure*}[t]
\centering
\includegraphics[width=7in]{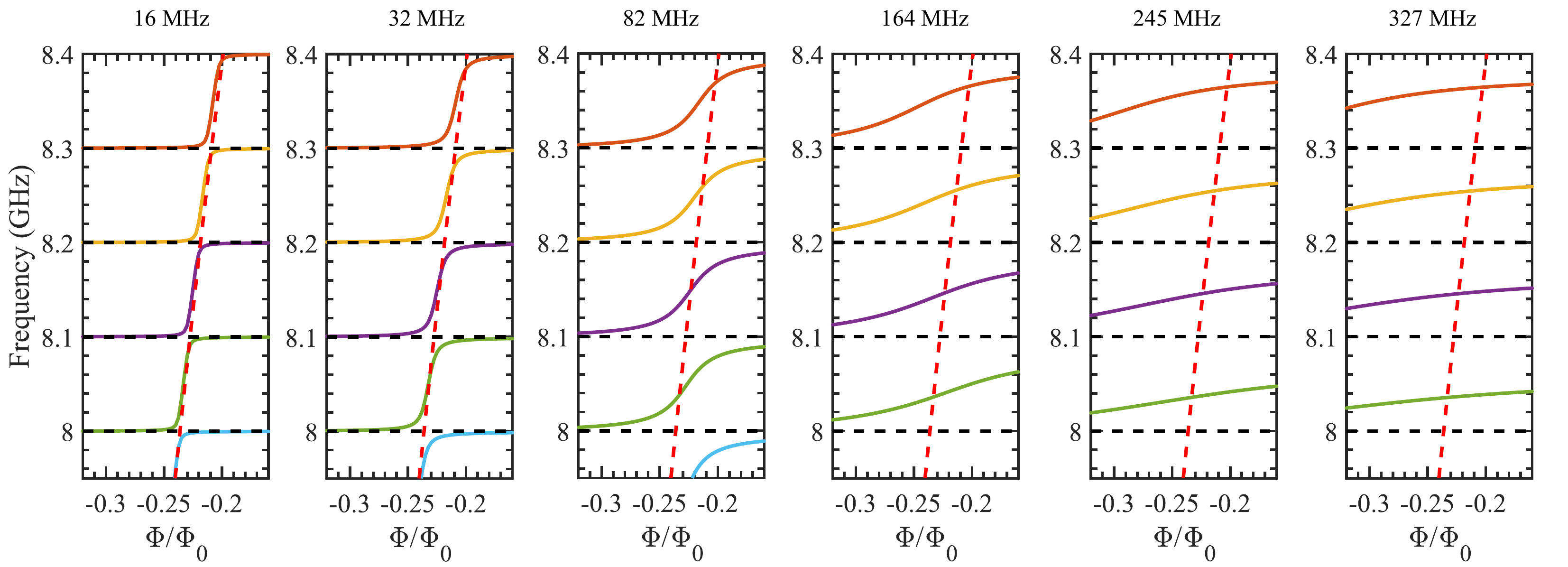}
  \caption{Numerical  solutions for a transmon coupled to four modes  with 100~MHz intermode spacing; bare mode frequencies are indicated by
horizontal dashed lines. Nearly vertical red dashed line corresponds to bare transmon frequency. Label at the top of each plot indicates the coupling strength $g_i/2\pi$ between each of the four modes and the transmon.    
\label{fig:couplingmat}}
\end{figure*}

With the ability to perform dispersive measurements of the qubit with the readout resonator (using $R_{in/out}$) while simultaneously driving a separate microwave signal to the hybrid metamaterial resonator (using $M_{in/out}$), we are able to observe the ac Stark shift of the qubit 0-1 transition \cite{Schuster2005stark} for different average numbers of photons in each of the metamaterial modes. With the qubit biased at 6.275~GHz, as indicated in Fig.~\ref{fig:Starkshifts}(g), we measure the qubit transition in spectroscopy while driving one of the six different metamaterial modes and scanning over 18~dB of power for the Stark drive to generate the plots in Fig.~\ref{fig:Starkshifts}(a-f). Additionally, for two of these metamaterial modes, we again perform qubit spectroscopy while scanning the frequency of the Stark drive for fixed power [Fig.~\ref{fig:Starkshifts}(h,i)]. 

%The observed shifts fit well for a model of a transmon qubit dispersively coupled to a single mode resonator. The Stark shift per photon experienced by the qubit is given by  $\chi=g_i^{2}\eta/ (\delta_i\eta-\delta_i^{2})$. Here  $\delta_i=\omega_q-\omega_i$, where $\omega_i$ is the metamaterial mode frequency and $\omega_q$ is the qubit's 0-1 transition frequency with no microwave drive to the metamaterial; $\eta$ is the qubit anhamornicity, defined as the difference between the 1-2 and 0-1 transition frequencies.  The observed shift of the qubit transition increases in magnitude with the power of the Stark drive when the frequency is resonant with a metamaterial mode, as with a qubit coupled to a single resonator~\cite{Schuster2005stark}. Moreover, if the metamaterial mode being driven falls between the qubit 0-1 and 1-2 transition frequencies, the straddling regime identified in Ref.~\cite{Koch2007}, we observe a change in the sign of the Stark shift.
%As explained in Appendix~\ref{app:stark}, the theoretical curves in Fig.~\ref{fig:Starkshifts}(a-f) were obtained with a single parameter fitting of the data from the two independent measurements of input power and frequency shift. This fit provides a mapping from the input power driving the hybrid metamaterial to the steady-state photon occupation number $\bar n$ in each mode, shown in Fig.~\ref{fig:Starkshifts}(j). 

The observed shift of the qubit transition increases in magnitude with the power of the Stark drive when the frequency is resonant with a metamaterial mode, as with a qubit coupled to a single resonator~\cite{Schuster2005stark}. Moreover, if the metamaterial mode being driven falls between the qubit 0-1 and 1-2 transition frequencies, the straddling regime identified in Ref.~\cite{Koch2007}, we observe a change in the sign of the Stark shift. The observed Stark shifts of the qubit with microwave driving of various metamaterial modes can be explained well by a model of a transmon qubit coupled dispersively to a single mode resonator, corresponding to the particular mode being driven (mode $i$), according to
%\begin{align}
%H/\hbar=\omega_i a^\dagger_i a_i +\left(\frac{\omega_{01}}{2}+\chi a^\dagger_i a_i\right)\sigma_z,
%\label{eq:Hamilstark}
%\end{align}

\begin{align}
H/\hbar=\omega_i \hat{\mathbb{M}_i} +\left(\frac{\omega_{q}}{2}+\chi \hat{\mathbb{M}_i}\right)\sigma_z,
\label{eq:Hamilstarkm}
\end{align}

where $\omega_i$ is the frequency of the relevant (single) metamaterial mode, $\hat{\mathbb{M}_i}=\sum_{m_i}m_i|m_i\rangle \langle m_i|$ is the number operator for mode $i$, $\omega_{q}$ is the qubit 0-1 transition frequency, $\chi=g_i^{2}\eta/ (\delta_i\eta-\delta_i^{2})$ is the qubit frequency shift per photon, $\delta_i=\omega_{q}-\omega_i$ is the detuning between the qubit and the metamaterial mode, and $\eta$ is the anharmonicity of the transmon, defined as the difference between the 1-2 and 0-1 transition frequencies.    The Stark tone at frequency $\omega_d$ drives the resonator to a coherent steady state with average photon number
\begin{align}
\bar n_i=\langle \hat{\mathbb{M}_i}\rangle=\frac{\Omega_i^2}{(\omega_i-\omega_d)^2+\frac{\kappa^2_i}{4}}.
\end{align}

Here, $\Omega_i$ is the effective drive amplitude for mode $i$ and  and $\kappa_i$ is the mode decay rate from separate measurements of the linewidth of each metamaterial mode (Fig.~\ref{fig:qvf}). Thus, $\bar n_i$ is proportional to the power delivered to the mode. Making a semiclassical approximation to the Hamiltonian in Eq. (\ref{eq:Hamilstarkm}), one finds the qubit Stark shift to be given by $\chi \bar n_i$. A single-parameter linear fit between the power measured at the chip for each mode frequency from a separate baseline cooldown and the observed Stark shift for each driven mode gives a map from the input drive power for each mode and the actual power delivered to the mode, $\Omega_i^2$. This fit parameter then allows us to compute the theory curves included in Fig. \ref{fig:Starkshifts}.

\section{Approaches for increasing qubit-metamaterial mode couplings}
\label{app:incg}
Although on our present device the coupling strength between the qubit and each metamaterial mode was always less than the spacing between modes $\Delta \omega_i$, in this section we consider the parameters for a hypothetical qubit-metamaterial device that could reach superstrong coupling \cite{Kuzmin2019}, where $g_i/\Delta \omega_i > 1$. In this regime, the qubit can be strongly coupled to multiple modes simultaneously. When combined with the ability to perform fast, non-adiabatic manipulation of the qubit transition frequency, this could be used to prepare multi-mode entangled states of photons over a range of metamaterial modes \cite{Egger13}. In this section, we consider various modifications to the metamaterial and qubit, some of which enhance $g_i$ and others that decrease $\Delta \omega_i$.

\begin{table}[b]\caption{Modified metamaterial parameters for hypothetical device with enhanced coupling strength used in AWR simulation. }% title of Table
\centering % used for centering table
\begin{tabular}{|c| c |c |c |}% centered columns (4 columns)
\hline\hline                        %inserts double horizontal lines
Label & Description & Value for & New value  \\ 
 &  &  measured chip &   \\[0.5ex]% inserts table %heading
\hline                  % inserts single horizontal line

$N_{l}$ & Number of LHTL unit cells & 42 & 82  \\ [0.5ex]% inserts table %heading
\hline 
$Z_{M}$ & Metamaterial impedance & 50 $\Omega$ & 200 $\Omega$ \\ [0.5ex]% inserts table %heading
\hline 
$N_{r}$ & Number of RHTL unit cells & N/A & 20  \\ [0.5ex]% inserts table %heading
\hline 

$L_{RH}$ & RHTL unit cell inductance & N/A & 0.35 nH  \\ [0.5ex]% inserts table %heading
\hline 
$C_{RH}$ & RHTL unit cell capacitance & N/A & 9.5 fF  \\ [0.5ex]% inserts table %heading
\hline 
$C_{QM}$ & Qubit-metametarial  & 4.3 fF & 50 fF\\
& coupling capacitance& & \\
\hline 
$C_{Q}$ & Qubit capacitance  & 48 fF & 50 fF\\

     % [1ex] adds vertical space
\hline%inserts single line

\end{tabular}\label{table:newpara}% is used to refer this table in the text
\end{table}

\begin{figure}[t]
\centering
\includegraphics[width=3.35in]{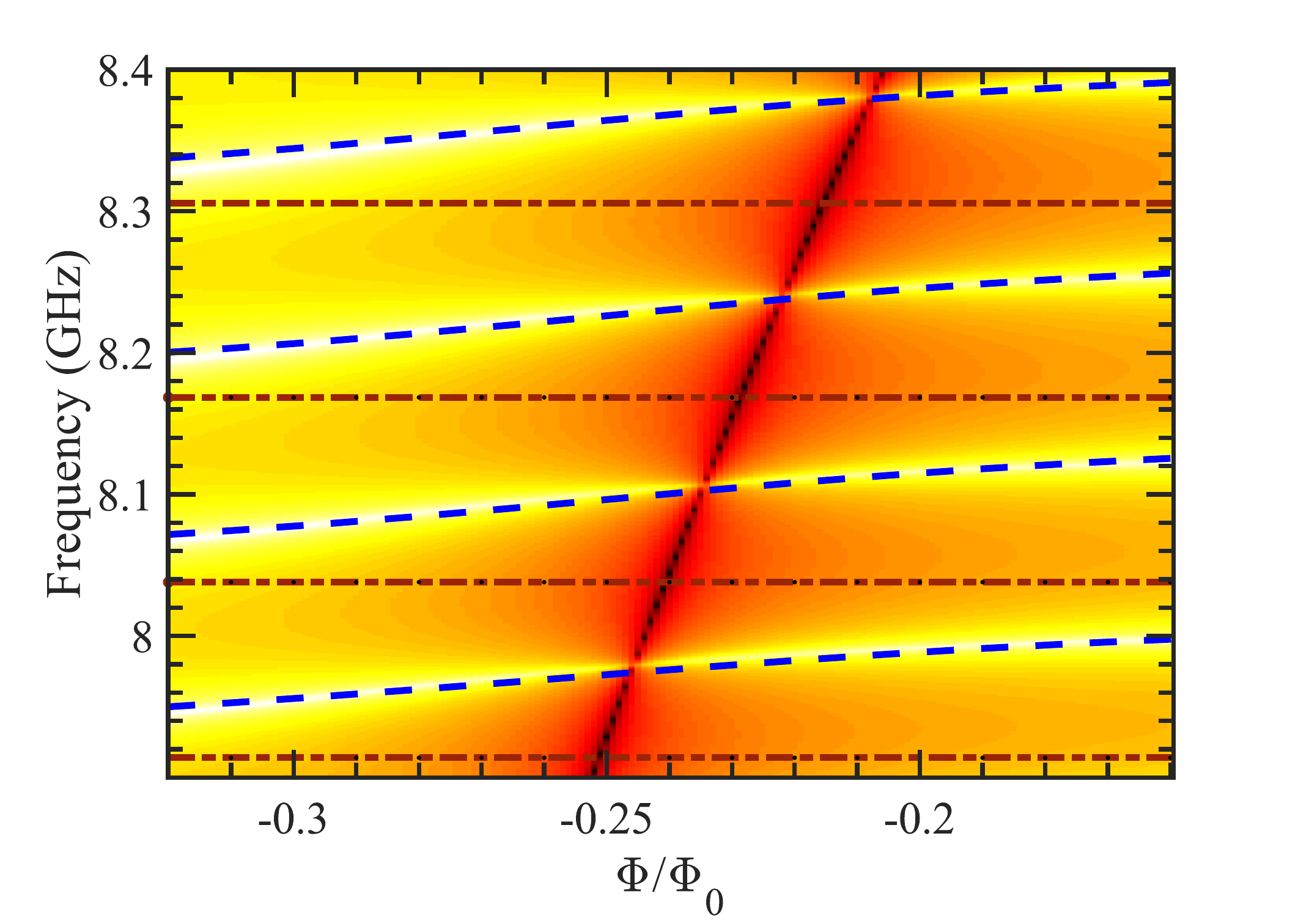}
  \caption{Simulation of hypothetical qubit-metamaterial device to achieve superstrong coupling through AWR circuit simulation of device described in text using parameters in Table \ref{table:newpara} and numerical solution to Hamiltonian (blue dashed lines) with adjusted coupling strength values to match features in AWR simulation. Black dotted line corresponds to bare qubit transition frequency. The brown  horizontal dashed-dotted lines show the bare frequencies for each of the four modes, 7.91~GHz, 8.04~GHz, 8.17~GHz, and 8.31~GHz, with extracted $g_{i}/2\pi$ values 220~MHz,  193~MHz,  180~MHz, and 178~MHz, respectively.
\label{fig:awrvsmat}}
\end{figure}

One approach for increasing $g_i$ involves increasing $C_{QM}$, as $g_i \propto C_{QM}/C_{\Sigma}^{1/2}$ [2]. We note that large coupling strengths between a qubit and resonator can also be achieved through inductive coupling \cite{Niemczyk2010}, but for now, we will restrict our design considerations to capacitive coupling. Another route for enhancing $g_i$ involves increasing the impedance of the metamaterial resonator. For the LHTL portion, this is straightforward to achieve by adjusting the values of $L_l$ and $C_l$. However, making a significant increase in the impedance for the RHTL portion is difficult with a CPW configuration. For example, a CPW impedance of 120~$\Omega$ requires a center conductor width of 500~nm and a 10-$\mu{\rm m}$ gap to the ground plane, which can be challenging to fabricate. As an alternative, we consider using a lumped-element implementation for the RHTL portion, which allows for larger impedances by choosing the inductor and capacitor values appropriately. We note that a lumped-element RHTL is a departure from our present design, but this should not introduce any complications since our device layout already includes chains of similar lumped-element inductors and capacitors for the LHTL and these would all get fabricated at the same time. In addition to increasing the coupling capacitance and resonator impedance, the length of the RHTL portion also impacts the coupling strength through the variation of the standing-wave amplitude at the location of the qubit. Besides increasing $g_i$, we can also decrease $\Delta \omega_i$ by adding more unit cells to the LHTL or by using a superlattice arrangement for the LHTL \cite{messinger2018left}, both of which increase the mode density since the number of modes between the IR and UV cutoff frequencies corresponds to the number of unit cells. Table~\ref{table:newpara} summarizes the various parameters for our hypothetical qubit-metamaterial device capable of achieving $g_i/\Delta \omega_i > 1$.

With the parameters of our hypothetical device described above, we characterize the coupling by simulating the microwave transmission through the metamaterial while tuning the qubit frequency. In order to build intuition about the splittings in the spectrum for the superstrong coupling regime, we also study the system Hamiltonian [presented in Appendix \ref{app:extraction}, Eq. (\ref{eq:hamil})] for various values of $g_i$. Because the size of the Hilbert space grows exponentially with the number of modes, in order to keep the numerical simulation tractable, we restrict the system to four modes spaced by 100~MHz  for the simulations shown in Fig.~\ref{fig:couplingmat}. For small $g_i$, such as the 16~MHz and 32~MHz plots, the solutions exhibit conventional vacuum Rabi splittings as the qubit passes through each of the individual metamaterial modes. For larger $g_i$, the splitting from one mode begins to merge with the splitting from the next mode, and by the point with 164~MHz couplings, the splittings become difficult to distinguish from the strongly shifted modes.

Figure \ref{fig:awrvsmat} contains a 500-MHz segment of the simulated transmission spectrum for our hypothetical device with the parameters from Table~\ref{table:newpara}  as a function of the qubit flux. We then run a numerical solution to the system Hamiltonian with four modes, corresponding to the bare mode frequencies in the frequency window of Fig.~\ref{fig:awrvsmat}, and adjust the coupling strengths $g_i$ in the Hamiltonian for these four modes to match the features in the AWR simulation. The blue dashed lines follow the Hamiltonian solutions and correspond to coupling strengths of between 178-220~MHz. Thus, $g_i/\Delta \omega_i$ ranges between 1.28 and 1.84, so that the hypothetical device with experimentally feasible parameters is capable of reaching the superstrong multimode coupling regime.

\verb||

\section{Conclusions}

%Novelty. Prospects for increasing $g$ and decresing mode spacing?
%
%Ability to operate qubit in bandgap below $\omega_{IR}/2\pi$ with reasonably long lifetime while remaining dispersively coupled to metamaterial modes; large Stark shifts...
%
%Achievable parameters and prospects for various analog sim...

In conclusion, we have demonstrated strong coupling of a transmon qubit to a metamaterial transmission line resonator. The left-handed dispersion of the structure resulted in a dense mode spectrum within the tuning window of the qubit transition frequency. We studied the influence of the metamaterial resonances on the qubit dynamics through measurements of energy relaxation and the ac Stark shift with different numbers of photons in several of the modes. As discussed in Sec. \ref{app:incg}, there is a straightforward path for a future device design that can enter the superstrong coupling regime \cite{Kuzmin2019}, where the coupling strength is larger than the mode spacing.

While the present sample was not configured for fast modulation of the qubit frequency, a future device could be designed with an on-chip flux-bias line for fast qubit tuning. This could enable new experiments with the qubit pulsed quickly between operating points below the metamaterial bandgap and resonance with different metamaterial modes \cite{Hofheinz2008}; alternatively, for such a device the qubit could be parametrically modulated at the sideband frequency between the qubit transition and a particular metamaterial mode \cite{Strand2013,naik2017random}. Both approaches could be used for swapping excitations between the qubit and the metamaterial. Such capabilities would allow for the preparation of complex multi-mode photonic states in the metamaterial that could be used for analog quantum simulations  with microwave photons \cite{Egger13,messinger2018left}, which could be made to interact through the nonlinearity coupled from the qubit when biased near the mode frequencies. Additionally, this system could be used for a quantum storage with microwave excitations in different metamaterial modes serving as memory elements.

\section{Acknowledgments}

We acknowledge useful discussions with Jaseung Ku and assistance with device fabrication from JJ Nelson. This work is supported by the U.S. Government under ARO Grants No. W911NF-14-1-0080 and No. W911NF-15-1-0248. Device fabrication is performed at the Cornell NanoScale Facility, a member of the National Nanotechnology Infrastructure Network, which is supported by the National Science Foundation (Grant NNCI-2025233). B.G.T. acknowledges support from CNPq INCT-IQ (465469/2014-0) and Coordenação de Aperfeiçoamento de Pessoal de Nível Superior - Brasil (CAPES) - Finance Code~001. 

\appendix
\section{Fabrication Details}
\label{app:fab}
The device was fabricated with two lithographic steps. Initially an 80-nm thick Nb film was sputter deposited onto a high-resistivity Si wafer. All of the circuit elements besides the qubit junctions were then patterned 
in a single photolithography step with UVN2300 deep-UV negative resist. After development, the pattern was etched using an inductively coupled plasma etcher with  BCl$_3$, Cl$_2$, and Ar. Next, the transmon junctions were defined with an electron-beam lithography step and a bilayer of MMA and PMMA to form the junction electrodes and airbridge for shadow evaporation. Following development in methyl isobutyl ketone, the junctions were formed by double-angle depostion of Al thin films with 35~nm for the first layer and 65~nm for the second layer. The tunnel barriers were formed by an in situ oxidation step in between the two film depositions; the unwanted Al was lifted off using dichloromethane following the vacuum step.

\begin{figure}[htb]
\centering
\includegraphics[width=3.35in]{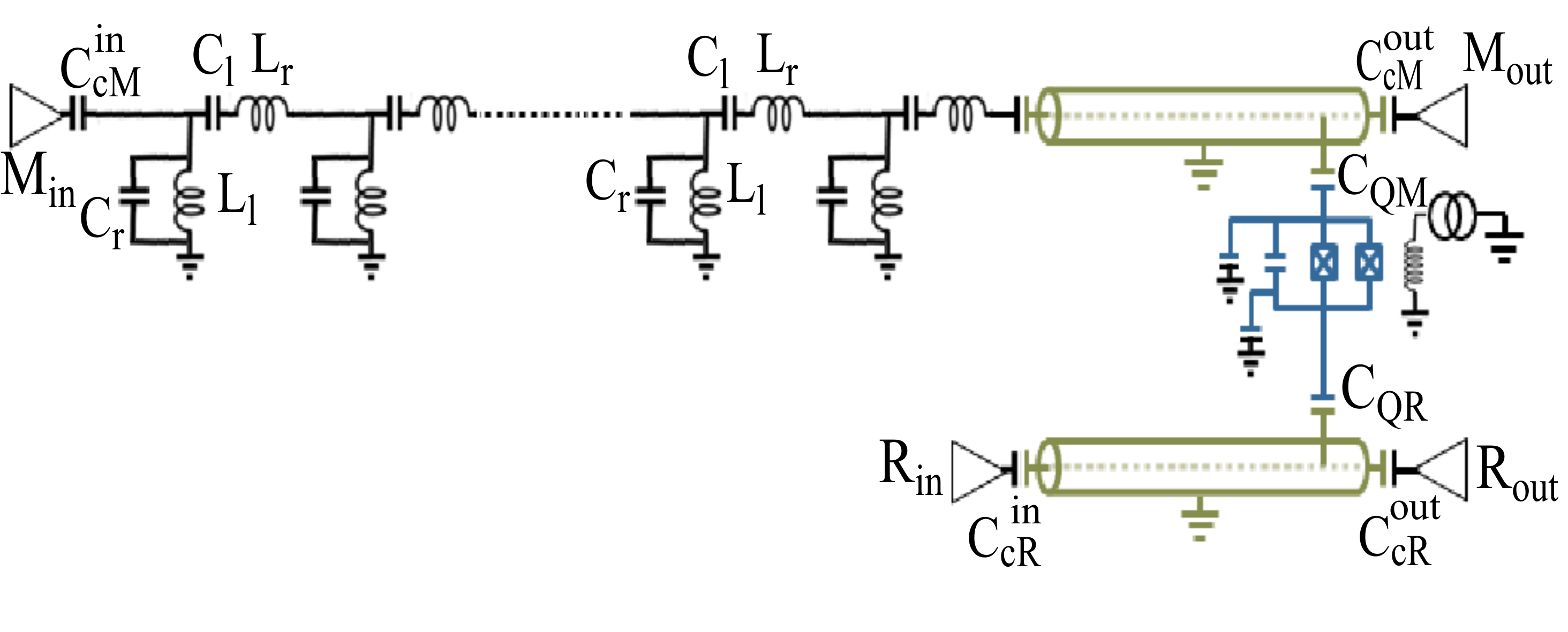}
  \caption{Circuit schematic of the device, including stray reactances.   
\label{fig:circuit}}
\end{figure}

\begin{table}[ht]\caption{Metamaterial parameters determined by finite element simulations of circuit layout.}% title of Table
\centering % used for centering table
\begin{tabular}{|c| c |c|}% centered columns (4 columns)
\hline\hline                        %inserts double horizontal lines
Label & Description & Value  \\ [0.5ex]% inserts table %heading
\hline                  % inserts single horizontal line
$L_{l}$ & Unit cell inductance & 0.7 nH  \\% inserting body of the table
\hline
$C_{l}$ & Unit cell capacitance & 250 fF   \\% inserting body of the table
\hline
$L_{r}$ & Unit cell stray inductance & 0.03 nH   \\% inserting body of the table
\hline
$C_{r}$ & Unit cell stray capacitance & 25 fF \\% inserting body of the table
\hline
$C_{cM}^{in}$ & Metamaterial input  & 30 fF \\
               & coupling capacitor &    \\
\hline
$C_{cM}^{out}$ & Metamaterial output  & 25 fF \\
               & coupling capacitor &    \\
     % [1ex] adds vertical space
\hline%inserts single line
\end{tabular}\label{table:meta}% is used to refer this table in the text
\end{table}

\section{Metamaterial and Qubit Parameters}
\label{app:para}
%The LHTL portion of the hybrid metamaterial resonator consisted of 42 unit cells of series interdigitated capacitors with 29 pairs of 4-$\mu$m wide/52-$\mu$m long fingers. Between each pair of capacitors was a meander-line inductor with 9 turns of 2-$\mu$m wide traces; the inductors were arranged in a staggered pattern with alternating inductors connected to the ground plane on either side of the LHTL, as in Ref.~\cite{Wang2019}. The input coupling capacitor to the LHTL portion was formed with a 4.9-$\mu$m wide gap to the input feedline. The RHTL portion of the hybrid metamaterial resonator was formed from a 5-mm long  CPW with a 10-$\mu$m wide center conductor and a 6-$\mu$m wide gap to the ground plane on either side. These parameters  allow us to match the impedance (about 50 Ohms)   between  the LHTL and RHTL segments. The output end of the RHTL consisted of an interdigitated capacitor for coupling to the output feedline. In order to account for stray reactances, we include parasitic series inductors $L_r$ and shunt capacitors $C_r$ in each unit cell (Fig. \ref{fig:circuit}) with parameters extracted from related structures in Ref.\cite{Wang2019} (Table \ref{table:meta}).

Table \ref{table:meta} lists the metamaterial parameters, which were determined by finite element simulations. In order to account for stray reactances, we include parasitic series inductors $L_r$ and shunt capacitors $C_r$ in each unit cell (Fig. 8) with parameters extracted from related structures that we modeled previously in Ref. \cite{Wang2019}.

\begin{figure}[htb]
\centering
\includegraphics[width=3.35in]{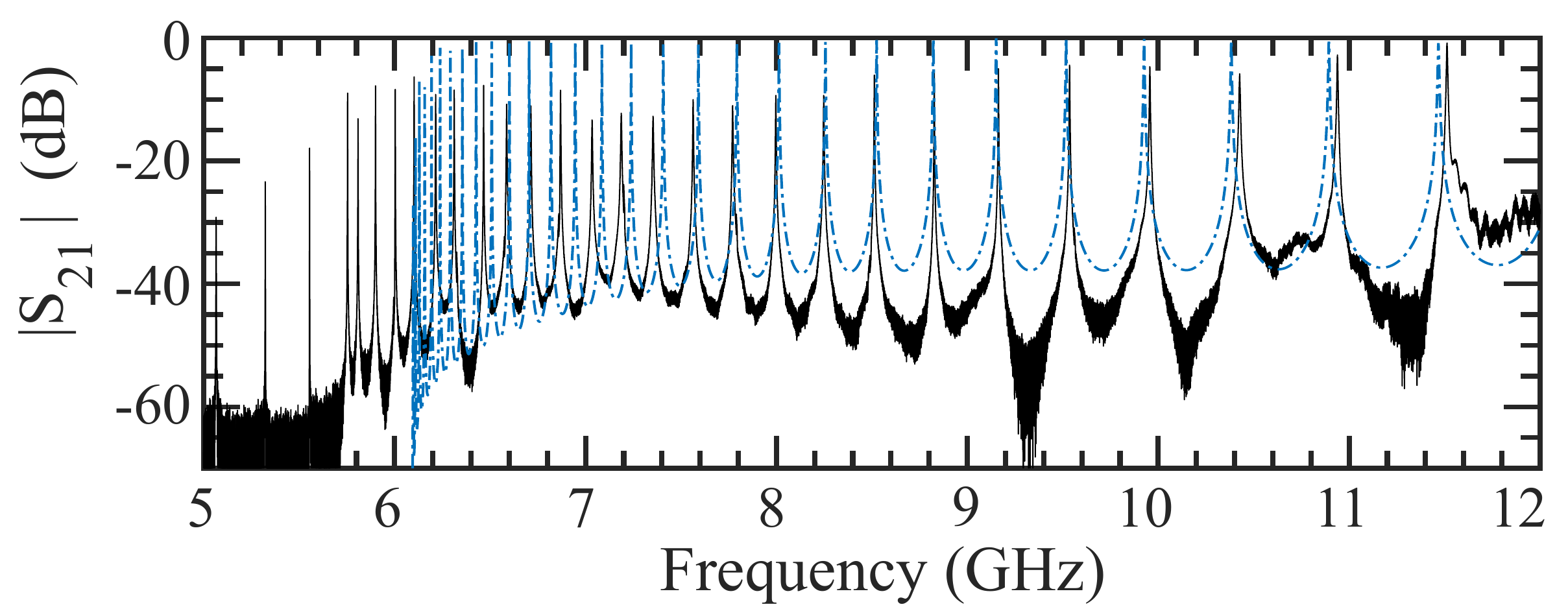}
  \caption{Measured transmission spectrum (black line) compared with spectrum from circuit simulation using AWR (blue dashed line).    
\label{fig:specawr}}
\end{figure}

\begin{figure}[t]
\centering
\includegraphics[width=3.35in]{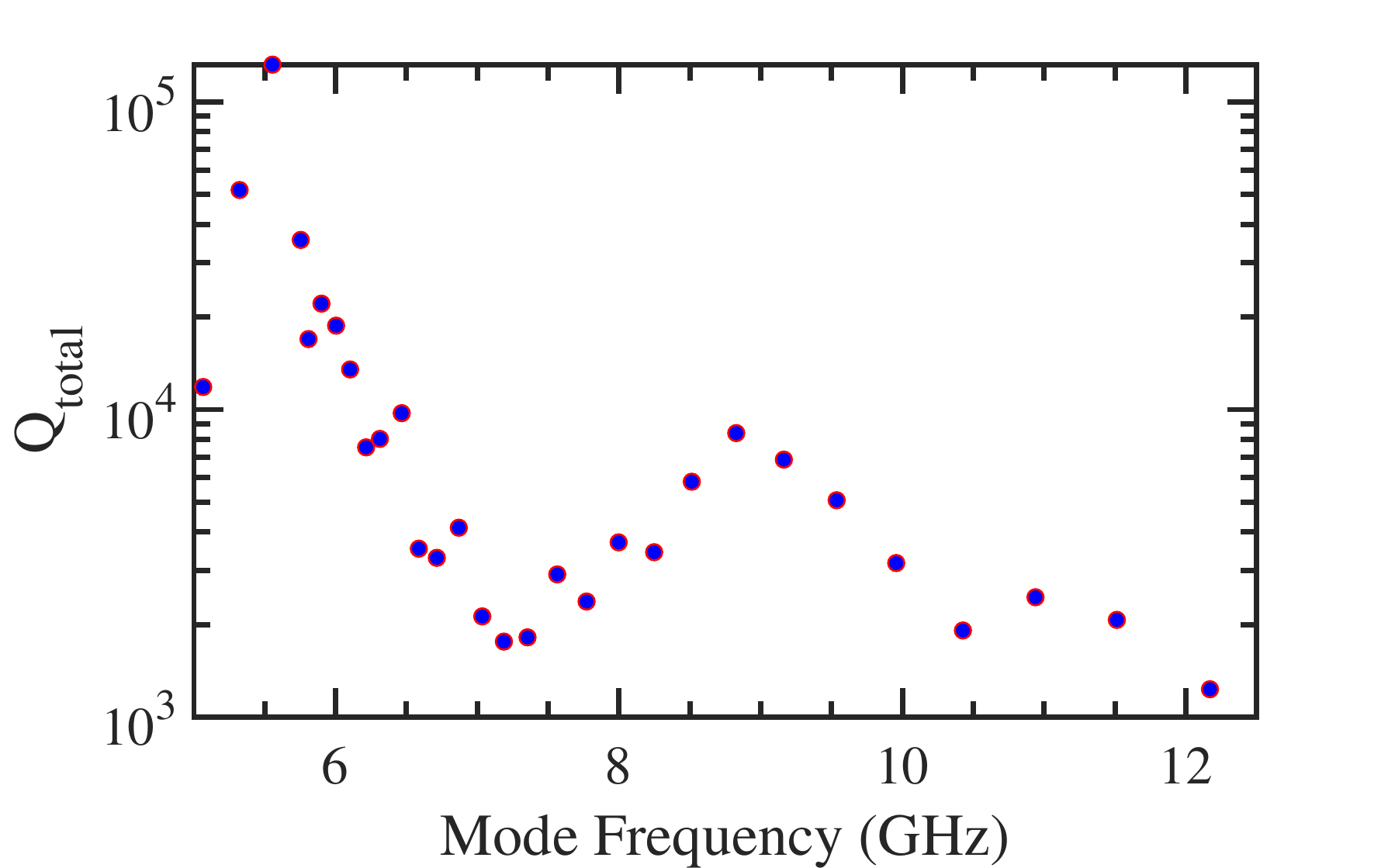}
  \caption{Total quality factor of  metamaterial modes  as function of frequency. The presence of the RHTL segment that is used to couple the metamaterial to the qubit results in the non-monotonic behavior of the mode quality factor vs. frequency through the variation in the standing wave pattern in the RHTL portion with frequency; by contrast, a simple LHTL resonator exhibits a monotonic decrease in quality factor for increasing frequency\cite{Wang2019}. 
\label{fig:qvf}}
\end{figure}

Figure \ref{fig:specawr} compares the measured transmission spectrum for the hybrid metamaterial with a circuit simulation of $S_{21}(f)$ using AWR Microwave Office with the parameters in the table. The spectra are reasonably close, with the most significant deviation at the low-frequency end, where the measured device has a softer infrared cutoff due to the staggered inductor configuration, as discussed in Ref.~\cite{Wang2019}. Figure \ref{fig:qvf} shows the total quality factor of modes extracted by Lorentzian fits to  measured spectrum  as function of frequency. Note that the quality factors for these modes are entirely dominated by coupling losses to external circuitry through the coupling capacitances.

Table \ref{table:sidecav} lists the qubit and readout resonator parameters, as determined by a combination of low-temperature measurements and circuit simulations.

\begin{table*}[ht]\caption{Qubit and readout resonator  parameters.}% title of Table
\centering % used for centering table
\begin{tabular}{|c| c |c| c|}% centered columns (4 columns)
\hline\hline                        %inserts double horizontal lines
Label & Description & Value & Method of Determination \\ [0.5ex]% inserts table %heading

\hline
$f_{01}^{max}$ & Maximum qubit frequency & 9.25 GHz & Qubit spectroscopy of the $f_{01}$ transition  \\
               &  & & at the flux-insensitive sweetspot \\
     % [1ex] adds vertical space
\hline%inserts single line

$C_{Q}$ &  Qubit shunt capacitance & 48 fF & Finite element simulations\\

\hline

                 % inserts single horizontal line
$g_{R}/2\pi$ & Qubit-readout resonator  & 65 MHz & Measurement of resonator  \\% inserting body of the table
               & coupling strength & & transmission $S_{21}$ vs. flux  \\
 \hline
 
$\omega_{R}/2\pi$ & Fundamental frequency of   & 7.07 GHz & Measurement of resonator   \\% inserting body of the table
               & readout resonator& & transmission $S_{21}$  \\
 \hline
 
$Q$ & Total quality factor of   & 15,463 & Measurement of resonator  \\% inserting body of the table
               &  readout resonator& &transmission $S_{21}$     \\
 \hline
$C_{cR}^{in}$ & Readout resonator input  & 1 fF & Finite element simulations  \\
               & coupling capacitor & &  \\
\hline
$C_{cR}^{out}$ & Readout resonator output  & 2 fF & Finite element simulations \\
               & coupling capacitor & &  \\
     % [1ex] adds vertical space
\hline 
$C_{QR}$ & Qubit-readout resonator  & 4.8 fF & Finite element simulations\\
               & coupling capacitor & &   \\
\hline
$C_{QM}$ & Qubit-metametarial  & 4.3 fF & Finite element simulations\\
               & coupling capacitor & &   \\
               
\hline
$C_{J}$ &  Junction capacitance  & 2.5 fF & Junction area from SEM image\\

     % [1ex] adds vertical space
\hline%inserts single line
 
\end{tabular}\label{table:sidecav}% is used to refer this table in the text
\end{table*}

\section{Measurement setup }
\label{app:measset}
The device was diced into a $6.25\times 6.25\,{\rm mm}^2$ chip and   wire-bonded into a machined aluminum sample box and covered by a lid with an integrated superconducting wirewound coil. The sample box was mounted on the cold finger of a  dilution refrigerator with a base temperature around 25~mK and was shielded by a single-layer cryogenic magnetic shield. The input coaxial lines to the metamaterial and readout resonator inputs each had 50~dB of cold attenuation as well as 12~GHz low-pass filters. The output lines also had 12~GHz low-pass filters before passing through a microwave relay switch, followed by a series of microwave isolators on the millikelvin stage. The output amplification consisted of a HEMT amplifier at 4~K plus 35~dB of room-temperature amplification. A room-temperature $\mu$-metal magnetic shield was used around the vacuum jacket of the dilution refrigerator.

\verb||

\section{Extraction of qubit-metamaterial mode couplings from spectra }
\label{app:extraction}
In order to determine the coupling strength $g_i$ between the transmon and  mode $i$ of the metamaterial with frequency $\omega_i$, we assume each mode can be represented as an independent harmonic oscillator coupled capacitively to the transmon. We then utilize standard circuit quantization \cite{Koch2007,Egger13} to derive the Hamiltonian for the transmon-coupled metamaterial, which, when written in the basis of transmon charge $|n\rangle$ and resonator excitation number $|m_i\rangle$ is given by

\begin{widetext}

\begin{equation}
\begin{split}
\hat{H}=\left[\sum_{n} \left(4 E_C (n-n_g)^2 |n\rangle \langle n| -\frac{E_J}{2}\left( |n+1\rangle \langle n|+|n\rangle \langle n+1|\right)\right)\right] \otimes \hat{\mathbb{I}} _{m}+\\
\sum_i \left[ \sum_{m_i} \hat{\mathbb{I}} _{n } \otimes \hbar \omega_i \left(|m_i\rangle \langle m_i|+\frac{1}{2}\right)+\sum_{n,m_i}\hbar g_in|n\rangle \langle n|\otimes \sqrt{m_i+1}(|m_i+1\rangle \langle m_i|+|m_i\rangle \langle m_i +1|) \otimes \hat{\mathbb{I}} _{m_{j\neq i} } \right].
\end{split}
\label{eq:hamil}
\end{equation}

\end{widetext}
Here, $\hat{\mathbb{I}} _{m}$ is the product of metatmaterial  mode identity operators, $\hat{\mathbb{I}} _{m_i}$ is the identity operator for metamaterial mode $i$, $\hat{\mathbb{I}}_n$ is the charge basis identity operator, $E_C=e^2/2C_{\Sigma}$ where $(C_{\Sigma}=C_Q+2C_J+C_{QR}+C_{QM})$ is the transmon charging energy, $n_g$ is the transmon polarization charge, and $E_J$ is the flux-tunable Josephson energy, which, for the case of symmetric junctions, is given by  

\begin{equation}
E_J=E_{J0}\left|\cos\left(\pi \frac{\Phi}{\Phi_0}\right)\right|,
\label{eq:Ej}
\end{equation}
where $E_{J0}$ is the maximum Josephson energy, $\Phi$ is the external flux applied to the transmon, and $\Phi_0 \equiv h/2e$ is the flux quantum. For our device, $E_{C}/h = 0.31$~GHz and $E_{J0}/h = 37$~GHz. Note that due to the large ratio of $E_J/E_C$ and the level of uniformity achieved in the Josephson junction fabrication, the dependence on $n_g$ and junction asymmetry are assumed to be negligible in the model for the fitting routine discussed below. 

Next, we implement a numerical minimization routine in Matlab to fit the lowest two eigenvalues of the model Hamiltonian to the frequency response of metamaterial transmission measurements versus flux $\Phi$ (main text, Fig. 2(a) and inset) in the vicinity of each vacuum Rabi splitting. For each fit, we truncate the transmon Hilbert space to 21 charge states and each resonator mode to 4 number states. In order to reduce the convergence time of the fit to an acceptable duration, for each vacuum-Rabi splitting $i$, we  include just a single metamaterial mode in the model (i.e.,  mode $i$ with frequency $\omega_i$).  Note that in each fit, $g_{i}$ is the only free parameter, with all other parameters in the Hamiltonian determined via independent measurements. 
In several trial cases, to estimate the error that results from neglecting the full spectrum of metamaterial modes, we performed fits where we solved the Hamiltonian including two nearest neighbor modes, rather than only one mode. We found the extracted value of each $g_i$ in these cases to be within 5\% of the value from the fits with only a single metamaterial mode. This is compatible with the coupling regime for our present device, where $g_i$ remains less than the intermode spacing $\Delta \omega_i$ between modes $i$ and $i+1$ over the range of our measurements..

\begin{figure}[htb]
\centering
\includegraphics[width=3.35in]{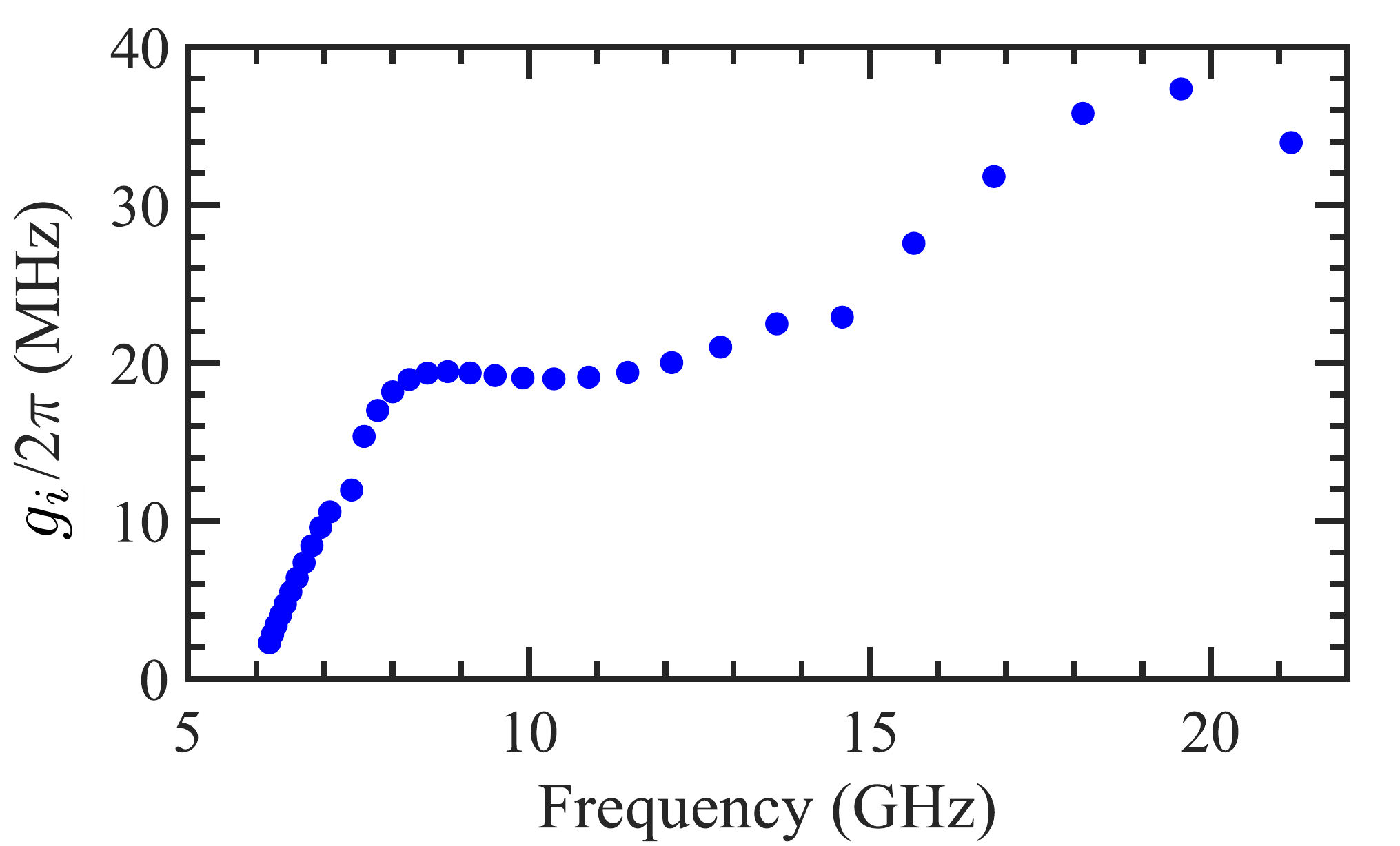}
  \caption{ Simulated $g_i$ from AWR/circuit model extended out to higher frequency for a hypothetical qubit with higher upper sweetspot. The slight deviation at 7.5 GHz and 15 GHz is  due to mode frequency approaching fundamental and first harmonic modes of readout resonator. 
\label{fig:awrg20GHz}}
\end{figure}

\section{Calculation of qubit-metamaterial mode couplings }
\label{app:gawr}
For modeling $g_i$, the coupling between metamaterial mode $i$ and the transmon qubit, we use a semi-classical approach involving the AWR Microwave Office circuit simulator \cite{AWR}. Note, all the circuit parameters used in the simulation are listed in Tables \ref{table:meta} and \ref{table:sidecav}. As shown in Fig. \ref{fig:circuit}, the LHTL section of the metamaterial resonator consists of 42 unit cells of series capacitors $C_l$ and inductors $L_l$ to ground, as in the measured device. In addition, we include parasitic effects in each cell consisting of a small lumped-element inductor $L_r$ in series with $C_l$ that accounts for the stray inductance of the interdigitated capacitor; a small lumped-element capacitor $C_r$ in parallel with $L_l$ accounts for the stray capacitance from the meander-line inductors. The values for these circuit parameters used in the simulations were chosen based on our earlier modeling of LHTL resonators \cite{Wang2019}, including simulations with Ansys Q3D \cite{Q3D} and Sonnet \cite{suitesversion}. We set the loss in the capacitors in AWR to correspond to an internal quality factor of $10^5$ \cite{Wang2019}.

For the AWR modeling, we treat the transmon as a tunable, lumped-element $LC$ oscillator, with $L$ being flux-tunable and the $L$ and $C$ values determined by the measured device. The coupling between the qubit and the RHTL CPW portion of the metamaterial resonator is set by the coupling capacitor $C_{QM}$, as listed in Table~\ref{table:sidecav}. The simulation also includes the separate CPW readout resonator (Fig. \ref{fig:circuit}), using parameters corresponding to the measured device, with coupling capacitor $C_{QR}$ to the qubit.

To extract the value of $g_i$, we simulate $S_{21}$ transmission measurements through the metamaterial (i.e., between $M_{in}$ and $M_{out}$ in Fig. \ref{fig:circuit}). We adjust $L$ for the qubit to simulate flux tuning and bring the qubit near resonance with mode $i$. We then scan the qubit $L$ to sweep it through resonance with mode $i$, and thus generate a simulated vacuum Rabi splitting. We thus determine the coupling strength to mode $i$ from the minimum spacing in the avoided-level crossing with the qubit, corresponding to twice the simulated coupling strength, $2g_i$. With this technique, we can simulate the frequency dependence of $g_i$ vs. mode frequency $\omega_i$, as in Fig. 2(c) of the main manuscript.

Unlike our measured device, where the upper sweet spot of the qubit is 9.25~GHz, the circuit simulations allow us to explore an artificial qubit with a much higher maximum frequency, so that we can study $g_i$ further along the metamaterial resonance spectrum. The plot in Fig. \ref{fig:awrg20GHz} exhibits multiple dips as the mode frequencies increase, which are due to the standing-wave structure in the RHTL portion. As the standing-wave pattern in the RHTL portion changes, the voltage level coupled to the qubit through $C_{QM}$, and hence the coupling strength $g_i$, can change.

\section{Calculation of Purcell loss}
\label{app:purcell}

The complex frequency dependence of the qubit lifetime that we observe in Fig. \ref{fig:T1} can be described by a combination of Purcell loss for a qubit coupled to a series of lossy resonant modes \cite{Houck2008} and dielectric loss with a frequency-independent loss tangent that is typically observed in frequency-tunable transmons \cite{Barends13, Hutchings17}:

\begin{equation}
\frac{1}{T_1}=\frac{1}{T_1^{Purcell}}+\frac{1}{T_1^{non-Purcell}},
\label{eq:Purcell1}
\end{equation}
where $T_1^{non-Purcell}(\omega)= A/\omega$  for some constant $A$. This,  of course, is a simplification, as even simple real devices with a single qubit coupled to a single resonator mode can exhibit structure in a measurement of $T_1(\omega)$ as the qubit passes through strongly coupled two-level system (TLS) or other spurious resonances at various frequencies \cite{Barends13}. Nonetheless, this gives us a starting point for modeling the frequency dependence of $T_1$ on our device.

Following the approach outlined in Ref.~\cite{Houck2008}, we model the multi-mode Purcell effect as

\begin{equation}
T_1^{Purcell}(\omega) = (C_{Q}+2C_{J})/{\rm Re}[Y(\omega)],
\label{eq:Purcell2}
\end{equation}
where $C_{Q}$ is the qubit shunt capacitance, $C_{J}$ is the single junction capacitance, and $Y(\omega)$ is the frequency-dependent complex admittance of the qubit environment. By modeling $Y(\omega)$ for our qubit environment and computing $T_1$ from Eqs. (\ref{eq:Purcell1}, \ref{eq:Purcell2}), we are able to compute the multi-mode Purcell loss curve in Fig. \ref{fig:T1}(e). The environment for our qubit consists of the  impedance of the readout resonator coupled to the qubit  $Z_R(\omega)$ and  the hybrid metamaterial resonator coupled to the qubit $Z_M(\omega)$:    $Y(\omega)=1/Z_R (\omega)+1/Z_M (\omega)$. From the coupling to the readout resonator, we have:

\begin{equation}
{Z_R(\omega)}=\frac{1}{i \omega C_{QR}}+\frac{1}{\frac{1}{Z_{RA}}+\frac{1}{Z_{RB}}},
\label{eq:read1}
\end{equation}
where $Z_{RA}$ and $Z_{RB}$ are the impedances of the two segments of the readout resonator on either side of the coupling element with length $l_A=$ 6.88~mm and $l_B=$ 0.792~mm, respectively, and given by the standard expression from Ref.~\cite{Pozar-book}:
\begin{equation}
Z_{RA}(\omega)=Z_0 \frac{Z_{LA}+i Z_0 \tanh(\gamma l_A)}{Z_0+i Z_{LA} \tanh(\gamma l_A)},
\label{eq:read2}
\end{equation}
where $Z_{0}=50\,\Omega$ is the characterstic impedance of the readout resonator transmission line, $Z_{LA}=1/i\omega C_{cR}^{in}+R_0$ and
$R_0$ is the source impedance on the input line, which is also 50~$\Omega$; $\gamma=\alpha+i\beta$ is the propagation constant, where $\alpha= 10^{-5}\pi/2l_A$ accounts for internal transmission line losses; $\beta=\omega\sqrt{\epsilon}/c$, where $\epsilon$ is the effective relative permittivity for the transmission line and $c$ is the speed of light. We get a similar expression for $Z_{RB}(\omega)$, except $Z_{LB}=1/i\omega C_{cR}^{out}+R_0$.

The impedance coupled to the qubit by the metamaterial resonator is given by
\begin{equation}
{Z_M(\omega)}=\frac{1}{i \omega C_{QM}}+\frac{1}{\frac{1}{Z_{MA}}+\frac{1}{Z_{MB}}},
\label{eq:lhl1}
\end{equation}
where  $Z_{MA}$ is the impedance of the RHTL segment of length 0.9 mm between the coupling point of the qubit along the RHTL and the output coupling capacitor of the metamaterial, which is described by a similar expression to Eq. (\ref{eq:read2}) with $Z_{L}=1/i\omega C_{cM}^{out}+R_0$; $Z_{MB}$ is the series impedance  of  the LHTL line and  RHTL segment of length 4~mm.

\begin{equation}
{Z_{MB}(\omega)}=Z_{0r} \frac{Z_{LHTL}+i Z_{0r} \tanh(\gamma l)}{Z_{0r}+i Z_{LHTL} \tanh(\gamma l)},
\label{eq:lhl1}
\end{equation}
where $Z_{0r}$ is again 50~$\Omega$ and $\gamma$ is the same as discussed previously following Eq.~(\ref{eq:read2}); $Z_{LHTL}$ represents the impedance of a LHTL with $N$ unit cells and unit cell length $\Delta x$, which we derived previously in  Ref.~\cite{Wang2019}: 

\begin{equation}
Z_{LHTL}=Z_{0l}\frac{e^{ik N\Delta x}+\Gamma e^{-ik N\Delta x}}{e^{-ik(-N+\frac{1}{2})\Delta x}-\Gamma  e^{ik(-N+\frac{1}{2})\Delta x}},\label{eqn:impedanceZN}
\end{equation}
with reflection coefficient
\begin{equation}
\Gamma=\frac{Z_se^{\frac{-ik\Delta x}{2}}-Z_{0l}}{Z_se^{\frac{ik\Delta x}{2}}+Z_{0l}}\label{eq:A20},
\end{equation} 
and input impedance
\begin{equation}
Z_{s}=1/i\omega C^{in}_{cM}+R_{0},
\label{eq:A12}
\end{equation}
where $R_0$ is the source impedance connected to the LHTL input; the characteristic impedance is given by

\begin{equation}
Z_{0l}=\left. \left(i\omega L_{r}+\frac{1}{i\omega C_{l}}\right)\middle/2i\sin(k\Delta x/2) \right., 
 \end{equation} 
and the wavenumber $k$ can be obtained from the LHTL dispersion relation
\begin{equation}
 k(\omega)\Delta x =
\cos^{-1}\left[1-\frac{1}{2}\left(\omega L_r-\frac{1}{\omega C_l} \right) \left(\omega C_r-\frac{1}{\omega L_l} \right) \right].
\label{eq:A11}
\end{equation}

\newpage 
\bibliographystyle{apsrev4-1-newv2}
\bibliography{meta-qubit-paper-Refv1}
\end{document}